\documentclass[prb,aps,onecolumn,amsmath]{revtex4}
\usepackage{graphicx}
\usepackage{dcolumn}
\usepackage{bm}
\begin{document}
\title{Dynamics of thin-film spin-flip transistors with perpendicular source-drain magnetizations}
\author{Xuhui Wang}
\author{Gerrit E. W. Bauer}
\affiliation{Kavli Institute of NanoScience, Delft University of Technology, 2628 CJ Delft,
The Netherlands}
\author{Axel Hoffmann}
\affiliation{Materials Science Division and Center for Nanoscale Materials, Argonne
National Laboratory, Argonne, Illinois 60439, USA}
\date{\today}

\begin{abstract}
A \textquotedblleft
spin-flip transistor\textquotedblright~is a lateral spin valve 
consisting of ferromagnetic source drain contacts to
a thin-film normal-metal island with an electrically floating
ferromagnetic base contact on top. We analyze the \emph{dc}-current-driven 
magnetization dynamics of spin-flip transistors in which the
source-drain contacts are magnetized perpendicularly to the device plane by
magnetoelectronic circuit theory and the macrospin Landau-Lifshitz-Gilbert
equation. Spin flip scattering and spin pumping effects are taken into
account. We find a steady-state rotation of the base
magnetization at GHz frequencies that is tuneable by the source-drain bias. We
discuss the advantages of the lateral structure for high-frequency generation
and actuation of nanomechanical systems over recently proposed
nanopillar structures.

\end{abstract}

\maketitle

\section{\label{sec:intro}Introduction}

Current induced magnetization excitation by spin-transfer
torque\cite{slonczewski,berger} attracts considerable attention because of
potential applications for magnetoelectronic devices. The prediction of
current-induced magnetization reversal has been confirmed experimentally in
multilayers structured into pillars of nanometer
dimensions.\cite{katine2000prl,myers2002prl,kiselevnature,rippardprl} The
devices typically consist of two ferromagnetic layers with a high (fixed
layer) and a low coercivity (free layer), separated by a normal metal spacer.
The applied current flows perpendicular to the interfaces. Often magnetic
anisotropies force the magnetizations into the plane of the magnetic layers.
Recently a number of theoretical proposals pointed out interesting dynamics
when the magnetization of one of the layers is oriented perpendicular to the
interface planes.\cite{kent,dieny,haiwenxi}

\begin{figure}[pb]
\includegraphics[scale=0.5]{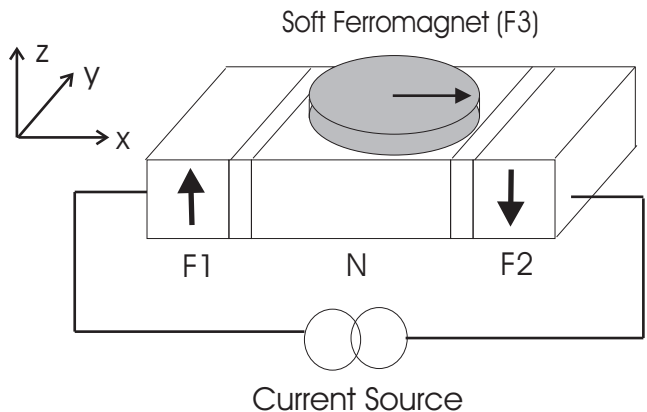}
\caption{The model system consists of hard-magnetic source and drain contacts
(F1 and F2) with antiparallel magnetizations perpendicular to the plane. On
the top of the normal metal N, a soft ferromagnetic film (F3) is deposited
with a slightly elliptical shape. The quantization direction, \textit{i.e.},
$z$-axis, is chosen parallel to the magnetization of the source and the
drain.}%
\label{fig:geometry}%
\end{figure}

Fundamental studies of charge and spin transport have also been carried out in
thin-film metallic conductors structured on top of a planar
substrate.\cite{jedema2001nature,jedema2002nature,vanwees,riken,tinkham,argonne}~%
The advantages compared to pillar structures are the flexible design and the
relative ease to fabricate multi-terminal structures with additional
functionalities such as the spin-torque transistor.\cite{bauer-spintorque} The
easy accessibility to microscopic imaging of the structure and magnetization
distribution should make the lateral structure especially suitable to study
current-induced magnetization dynamics. Previous studies focused on the static
(dc) charge transport properties, but investigations of the dynamics of
laterally structured devices are underway.\cite{tatara,vanwees-lateral}
Recently, non-local magnetization switching in a lateral spin valve structure
has been demonstrated.\cite{kimura-condmat2005} In the
present paper we investigate theoretically the dynamics of a lateral spin
valve consisting of a normal metal film that is contacted by two magnetically
hard ferromagnets. As sketched in Fig.~\ref{fig:geometry}, a (nearly) circular
and magnetically soft ferromagnetic film is assumed deposited on top of the
normal metal to form a spin-flip transistor.\cite{brataasprl} We concentrate
on a configuration in which the magnetization direction of the source-drain
contacts lies perpendicular to the plane of the magnetization of the third
(free) layer. This can be realized either by making the contacts from a
material that has a strong crystalline magnetic anisotropy forcing the
magnetization out of the plane, such as Co/Pt multilayers,\cite{copt} or by
growing the source/drain ferromagnetic contacts into deeply etched groves to
realize a suitable aspect ratio. In such a geometry, the magnetization of the
free layer precesses around the demagnetizing field that arises when the
magnetization is forced out of the plane by the spin-transfer torque, as has
been discussed in Refs.~ \onlinecite{kent,dieny,haiwenxi}. Therefore, as long
as the out-of-plane magnetization of the free layer remains small, the free
layer magnetization will always stay almost perpendicular to the source and
drain magnetizations. In the present article we analyze in depth the coupled
charge-spin-magnetization dynamics in such current-biased thin-film
\textquotedblleft magnetic fans\textquotedblright\ and point out the
differences and advantages compared to the perpendicular pillar structures. A
convenient and accurate tool to compute the dynamic properties of our device
is the magnetoelectronic circuit theory for charge and spin transport
\cite{brataasprl} coupled to the Landau-Lifshitz-Gilbert equation in the
macrospin model. We include spin flip scattering in normal and ferromagnetic
metals and the spin-pumping effect.
\cite{yaroslavprl,yaroslavrmp}

The article is organized as follows: In Section \ref{sec:formalism}, we
briefly review the Landau-Lifshitz-Gilbert equation including the current
driven and spin-pumping torques that can be derived by circuit theory. In
Section \ref{sec:dyna}, the specific results for our ``magnetic fan" are
presented. The potential applications will be discussed in Section
\ref{sec:app}. Section \ref{sec:concl} is devoted to the conclusion.

\section{\label{sec:formalism} Formalism}

We are interested in the magnetization dynamics of the soft ferromagnetic
island (i.e., composed of permalloy) on top of the normal film as sketched in
the Fig.~\ref{fig:geometry}. The Landau-Lifshitz-Gilbert (LLG) equation in the
macro-spin model, in which the ferromagnetic order parameter is described by a
single vector $\mathbf{M}$ with constant modulus $M_{s}$, appears to
describe experiments of current-driven magnetization dynamics
well,\cite{krivorotov} although some open questions remain.\cite{xiao} 
Micromagnetic calculations of the perpendicular
magnetization configuration in the pillar structure suggest a steady
precession of the magnetization.\cite{dieny} The LLG equation for isolated
ferromagnets has to be augmented by the magnetization torque $\mathbf{L}$ that
is induced by the spin accumulation in proximity of the interface as well as
the spin pumping:
\begin{equation}
\frac{1}{\gamma}\frac{d\mathbf{m}}{dt}=-\mathbf{m}\times\mathbf{H}_{eff}%
+\frac{\alpha_{0}}{\gamma}\mathbf{m}\times\frac{d\mathbf{m}}{dt}+\frac
{1}{VM_{s}}\mathbf{L} \label{LLG}%
\end{equation}
where $\gamma$ is the gyromagnetic constant, $\mathbf{m}=\mathbf{M}/M_{s}$ and
$\mathbf{H}_{eff}$ is the magnetic field including demagnetizing, anisotropy
or other external fields. $\alpha_{0}$ is the Gilbert damping constant and $V$
is the volume of the isolated bulk magnet.
\[
\mathbf{L}=-\mathbf{m}\times\left(  \mathbf{I}_{s}^{(p)}+\mathbf{I}_{s}%
^{(b)}\right)  \times\mathbf{m},
\]
where $\mathbf{I}_{s}^{(p)}$ and $\mathbf{I}_{s}^{(b)}$ denote the
pumped\cite{yaroslavprl} and bias-driven\cite{slonczewski,berger} spin
currents leaving the ferromagnet, respectively, and the vector products
project out the components of the spin current normal to the magnetization direction.

In magnetoelectronic circuit theory a given device or circuit is split into
nodes and resistors. In each node a charge potential and spin accumulation is
excited by a voltage or current bias over the entire device that is connected
to reservoirs at thermal equilibrium or by spin pumping. The currents are
proportional to the chemical potential and spin accumulation differences over
the resistors that connect the island to the nodes. The Kirchhoff rules
representing spin and charge conservation close the system of equations that
govern the transport. In the following we assume that the ferromagnetic layer
thickness is larger than the magnetic coherence length $\lambda_{c}
=\pi/\left\vert k_{F}^{\uparrow}-k_{F}^{\downarrow}\right\vert $ in terms of
the majority and minority Fermi wave numbers that in transition metal
ferromagnets is of the order of \AA ngstr\"oms.

Let us consider a ferromagnet-normal metal ($F|N$) interface in which the
ferromagnet is at a chemical potential $\mu_{0}^{F}$ and spin accumulation
$\mu_{s}^{F}\mathbf{m}$\ (with magnetization direction $\mathbf{m}%
$),\textbf{\ }whereas the normal metal is at $\mu_{0}^{N}$ and spin
accumulation $\mathbf{s}$. The charge current (in units of Ampere) and spin
currents (in units of Joule), into the normal metal are \cite{brataaseur}
\begin{align}
I_{c}  &  =\frac{e}{2h}[2g(\mu_{0}^{F}-\mu_{0}^{N})+pg\mu_{s}^{F}-pg\mathbf{m}%
\cdot\mathbf{s}]\\
\mathbf{I}_{s}^{\left(  b\right)  }  &  =\frac{g}{8\pi}[2p(\mu_{0}^{F}-\mu
_{0}^{N})+\mu_{s}^{F}-(1-\eta_{r})\mathbf{m}\cdot\mathbf{s}]\mathbf{m}%
\nonumber\\
&  -\frac{g}{8\pi}\eta_{r}\mathbf{s}-\frac{g}{8\pi}\eta_{i}(\mathbf{s}%
\times\mathbf{m}) \label{eq:current}%
\end{align}
where $\mu_{0}^{F}$ and $\mu_{0}^{N}$ are the chemical potentials in the
ferromagnets and normal metal, respectively. $g^{\uparrow},$ $g^{\downarrow}$
are the dimensionless spin dependent conductances with polarization
$p=(g^{\uparrow}-g^{\downarrow})/(g^{\uparrow}+g^{\downarrow})$ and total
contact conductance $g=g^{\uparrow}+g^{\downarrow}$. In the
Landauer-B\"{u}ttiker formalism
\begin{equation}
g^{\uparrow(\downarrow)}=M-\sum_{nm}|r_{\uparrow(\downarrow)}^{nm}|^{2}%
\end{equation}
where $M$ is the total number of channels and $r_{\uparrow(\downarrow)}^{nm}$
is the reflection coefficient from mode $m$ to mode $n$ for spin up(down)
electrons. The spin transfer torque is governed by the complex spin-mixing
conductance $g^{\uparrow\downarrow}$, given by\cite{brataaseur}
\begin{equation}
g^{\uparrow\downarrow}=M-\sum_{nm}r_{\uparrow}^{nm}(r_{\downarrow}^{nm}%
)^{\ast}~,
\end{equation}
introduced in Eq. (\ref{eq:current}) in terms of its real and imaginary part
as $\eta_{r}=2\text{Re}g^{\uparrow\downarrow}/g$ and $\eta_{i}=2\text{Im}%
g^{\uparrow\downarrow}/g$. All conductance parameters can be computed from
first principles as well as fitted to experiments.

Slonczewski's spin transfer torque can then be written as
\begin{equation}
-\mathbf{m}\times\mathbf{I}_{s}^{\left(  b\right)  }\times\mathbf{m}=\frac
{g}{8\pi}\eta_{r}[\mathbf{s}-(\mathbf{s}\cdot\mathbf{m})\mathbf{m}]+\frac
{g}{8\pi}\eta_{i}(\mathbf{s}\times\mathbf{m}) . \label{eq:perpspincur}%
\end{equation}
The spin-pumping current is given by\cite{yaroslavprl}
\begin{equation}
\mathbf{I}_{s}^{\left(  p\right)  }=\frac{\hbar}{8\pi}g\left(  \eta
_{r}\mathbf{m}\times\frac{d\mathbf{m}}{dt}+\eta_{i}\frac{d\mathbf{m}}%
{dt}\right)~.
\end{equation}

We consider for simplicity the regime in which the spin-flip diffusion length
$l_{sf}^{N}$ in the normal metal node is larger than the size of the normal
metal region.\cite{vanwees} Charge and spin currents into the normal metal
node are then conserved such that\cite{brataasprl}
\begin{align}
\sum_{i}I_{c,i}  &  =0\label{ccon}\\
\sum_{i}\left(  \mathbf{I}_{s,i}^{(p)}+\mathbf{I}_{s,i}^{(b)}\right)   &
=\mathbf{I}_{s}^{sf}. \label{scon}%
\end{align}
where we introduce a leakage current due to the spin-flip scattering
$\mathbf{I}_{s}^{sf}=g_{sf}\mathbf{s}/{4\pi}$ and
$g_{sf}={h\nu_{DOS}\text{V}_{N}}/{\tau_{sf}^{N}}$
is the conductance due to spin flip
scattering, where $\nu_{DOS}$ is the (on-spin)density of state of the electrons in the
normal metal, $\tau_{sf}^{N}$ is the spin flip relaxation time and $V_{N}$ the
volume of the normal metal node.

The polarization of the source-drain contacts is supposed to be an
effective one including the magnetically active region of the bulk
ferromagnet with thickness governed by the spin-flip diffusion
length in the ferromagnet. For the free magnetic layer $F3$, the
perpendicular component of the spin current is absorbed to generate
the spin transfer torque. The collinear current has to fulfill the
boundary conditions in terms of the chemical potential
$\mu_{s}^{F}=\mu_{\uparrow}-\mu_{\downarrow}$
governed by the diffusion equation%
\begin{equation}
\frac{\partial^{2}\mu_{s}^{F}(z)}{\partial z^{2}}=\frac{\mu_{s}^{F}%
(z)}{\left(  l_{sd}^{F}\right)  ^{2}}.
\end{equation}
where $l_{sd}^{F}$ is the spin flip diffusion length in the ferromagnet.

\section{\label{sec:dyna}Spin transfer torque and magnetic fan effect}

In this Section, we solve the Landau-Lifshitz-Gilbert equation including
expressions for the spin-transfer torque on the free layer according to the
circuit theory sketched above.

\subsection{Currents and spin torque}

In metallic structures the imaginary part of the mixing conductance
is usually very small and may be disregarded, \textit{i.e.},
$\eta_{i}\simeq0$. The source and drain contacts $F1|N$ and $F2|N$
are taken to be identical: $g_{1}=g_{2}=g$, $p_{1}=p_{2}=p$ and
$\eta_{r1}=\eta_{r2}\equiv\eta_{r}$. For $F3|N$ we take
$\eta_{r3}\equiv\eta_{3}$. In our device, the directions of the
magnetization of the fixed magnetic leads are
$\mathbf{m}_{1}=(0,0,1)$ and $\mathbf{m}_{2}=(0,0,-1)$. For the free
layer we allow the magnetization
$\mathbf{m}_{3}=(m_{x},m_{y},m_{z})$ to be arbitrary. We assume that
$F3$ is a floating contact in which the the chemical potential
$\mu_{0}^{F3}$ adjusts itself such that the net charge current
through the interface $F3|N$ vanishes:
\begin{equation}
I_{c}^{(3)}=\frac{eg_{3}}{2h}[2(\mu_{0}^{F3}-\mu_{0}^{N})+p_{3}\mu_{s}%
^{F3}-p_{3}\mathbf{s}\cdot\mathbf{m}_{3}]=0.
\end{equation}
Applying a bias current $I_{0}$ on the two ferromagnetic leads, $F1$
and $F2$, the conservation of charge current in the normal metal
then gives $I_{c}^{(1)}=-I_{c}^{(2)}=I_{0}$. At the $F3|N$
interface, the continuity of the longitudinal spin current dictates
\begin{equation}
\sigma_{\uparrow}\left(\frac{\partial\mu_{\uparrow}}{\partial
z}\right)_{z=0}
-\sigma_{\downarrow}\left(\frac{\partial\mu_{\downarrow}}{\partial
z}\right)_{z=0} =\frac{2e^{2}}{\hbar
A}\mathbf{I}_{s,3}\cdot\mathbf{m}_{3}
\end{equation}
where $\sigma_{\uparrow}$($\sigma_{\downarrow}$) is the bulk
conductivities of spin up (down) electrons in the ferromagnet and
$A$ the area of the interface. Choosing the origin of the $z$ axis
is at the $F3|N$ interface and assuming $F3$ to be of thickness $d$,
\begin{equation}
\sigma_{\uparrow}\left(\frac{\partial\mu_{\uparrow}}{\partial
z}\right)_{z=d}
-\sigma_{\downarrow}\left(\frac{\partial\mu_{\downarrow}}{\partial
z}\right)_{z=d} =0~.
\end{equation}
With both boundary conditions, the diffusion equation can be solved
for the spin accumulation in $F3$
\begin{equation}
\mu_{s}^{F}(z)=\frac{\zeta_{3}\cosh(\frac{z-d}{l_{sd}^{F}})\mathbf{s}\cdot\mathbf{m}_{3}}
{\left[\zeta_{3}+\tilde{\sigma}\tanh(\frac{d}{l_{sd}^{F}})\right]\cosh(\frac{d}{l_{sd}^{F}})}
\label{saf}
\end{equation}
where $\zeta_{3}=g_{3}(1-p_{3}^{2})/4$ characterizes the contact
$\text{F3}|\text{N}$ and $\tilde{\sigma}=h
A\sigma_{\uparrow}\sigma_{\downarrow}/(e^{2}l_{sd}^{F}(\sigma_{\uparrow}+\sigma_{\downarrow}))$
describes the bulk conduction properties of the free layer with
arbitrary $\mathbf{m}_{3}$. The limit $d\ll l_{sd}^{F}$ corresponds
to negligibly small spin-flip, which implies
$\tanh{(d/l_{sd}^{F})}\simeq0$. Near the interface, the spin
accumulation in $F3$ then reduces to
\begin{equation}
\mu_{s}^{F3}=\mathbf{s}\cdot\mathbf{m}_{3}~.
\label{saf0}%
\end{equation}
In this limit, $\mathbf{I}_{s}^{(3)}\cdot\mathbf{m}_{3}=0$ the
collinear component of the spin current vanishes.

By solving the linear equations generated by Eqs. (\ref{ccon},\ref{scon}), we
obtain the spin accumulation $\mathbf{s}$ in the normal metal node,
\begin{equation}
\mathbf{s}=\hat{\mathbf{C}}\cdot
[8\pi\mathbf{I}_{s}^{(p)}+\mathbf{W}_{b}]
\label{eq:spin-accu}%
\end{equation}
where the elements of the symmetric matrix $\hat{\mathbf{C}}$ are
given in Appendix \ref{sec:appa} and
$\mathbf{W}_{b}=(0,0,2phI_{0}/e)$ is a bias-vector. Eq.
(\ref{eq:spin-accu}) contains contribution due to bias current and
spin pumping effect. The spin accumulation in the ferromagnet
Eq.~(\ref{saf}) should be substituted in Eq.~(\ref{eq:spin-accu}) to
give the spin accumulation in the normal metal, from which the spin
transfer torque can be determined according to Eq.
(\ref{eq:perpspincur}). For an ultrathin film, the spin transfer
torque, including pumping effect and spin accumulation in the
ferromagnet, reads,
\begin{equation}
\mathbf{L}=\frac{\eta_{3}g_{3}}{8\pi} \hat{\mathbf{\Pi}} \cdot
[8\pi\mathbf{I}_{s}^{(p)}+\mathbf{W}_{b}]~,
\label{eq:spin-torque}
\end{equation}
with the elements of $\mathbf{\Pi}$ listed in Appendix.

\subsection{Dynamics of the free layer}

After the bias current is switched on, a spin accumulation builds up in
the normal metal. At the beginning, the spin-transfer torque exerted
on the magnetization of the free layer ($\text{F}3$) causes a
precession out of the plane, hence generating a demagnetizing field
$\mathbf{H}_{A}$ that is oriented perpendicular to the film plane.
Subsequently the magnetization precesses around $\mathbf{H}_{A}$ and
as long as the current $I_{0}$ continues, the rotation persists. In
order to determine the dynamics of the magnetization, we apply the
spin torque term $\mathbf{L}$ [Eq.~(\ref{eq:spin-torque})] to the
Landau-Lifshitz-Gilbert (LLG) equation (\ref{LLG}). Crystalline
anisotropies in $\text{F}3$ may be disregarded for soft ferromagnets
such as permalloy. The effective field in the LLG equation then
reduces to
\begin{equation}
\mathbf{H}_{A}=-\mu_{0}M_{s}(N_{x}m_{x},N_{y}m_{y},N_{z}m_{z})~,
\end{equation}
where $N_{x}$, $N_{y}$ and $N_{z}$ are the demagnetizing factors
determined by the shape of the film.\cite{osborn} The anisotropy
field keeps the magnetization in the plane when the torque is zero.
The spin torque generated by the current bias forces the
magnetization out of plane, hence triggering the nearly in-plane
rotation of the magnetization. Substituting the spin-torque term Eq.
(\ref{eq:spin-torque}) into Eq. (\ref{LLG}), we obtain for the
following LLG equation,
\begin{widetext}
\begin{equation}
\frac{1}{\gamma}\frac{d\mathbf{m}}{dt}=-\mathbf{m}\times\mathbf{H}_{A}%
+\frac{1}{\gamma}\left(\alpha_{0}+\overleftrightarrow{\alpha}^{\prime}\right)\mathbf{m}\times\frac{d\mathbf{m}}{dt}
+\mathbf{H}_{st}(I_{0})
\label{eq:full-llg}
\end{equation}
\end{widetext}Here the last vector
\begin{equation}
\mathbf{H}_{st}(I_{0})=\frac{\hbar}{2e}\Lambda_{st}\frac{I_{0}}{M_{s}%
V}(-m_{x}m_{z},-m_{y}m_{z},1-m_{z}^{2})~. \label{eq:effield}
\end{equation}
is the effective field induced by the spin-transfer torque that depends 
on the position of the magnetization and the device parameter
\begin{equation}
\Lambda_{st}=\frac{p\eta_{3}g_{3}\mathcal{G}_{1}}{\mathcal{G}_{t}\mathcal{G}_{3}+2(p^{2}-1+\eta)g\mathcal{G}_{4}(1-m_{z}^{2})},
\label{eq:lambda}%
\end{equation}
where $\mathcal{G}_{i}$'s are introduced in Appendix A. According to
Eq.~(\ref{eq:lambda}), we can accurately engineer the device
performance by tuning the conductances and polarizations. Compared
with the original LLG equation, a new dimensionless parameter
entering the calculation
\begin{equation}
\overleftrightarrow{\alpha}^{\prime}=\frac{\gamma\hbar
(\text{Re}g^{\uparrow\downarrow})^{2}}{2\pi V
M_{s}}\hat{\mathbf{\Pi}}
\end{equation}
reflects the tensor character of the pumping-induced additional
Gilbert damping.\cite{xuhuiwangunpub} Choosing contact $F3|N$ to be
metallic and the others to be tunneling barriers, the condition
$g_{3}\gg g,g_{sf}$ can be realized. In that limit
$\overleftrightarrow{\alpha}^{\prime}$ reduces to
\begin{equation}
\alpha^{\prime}=\frac{\gamma\hbar}{4\pi VM_{s}}\operatorname{Re}%
g_{3}^{\uparrow\downarrow}~,
\end{equation}
which agrees with the enhanced Gilbert damping derived in Ref.
~\onlinecite{yaroslavprl}. In the following, we take $\alpha=\alpha_{0}%
+\alpha^{\prime}$ to be the enhanced Gilbert damping constant.

\subsubsection{Vanishing in-plane anisotropy}

Here we rewrite the free layer magnetization in two polar angles $\phi$ (in-plane)
and $\theta$ (out-of plane) such that 
$\mathbf{m}=(\cos\theta\cos\phi,\cos\theta\sin\phi,\sin\theta)$ and assuming 
a small $z$-component, i.e., $m_{z}=\sin\theta\approx \theta$ and $\cos\theta\approx 1$.
When the free layer is a round flat disk with demagnetizing factors
$N_{x}=N_{y}\approx 0$ and $N_{z}\approx 1$, the Eqs. (\ref{eq:full-llg}) reduce to:
\begin{align}
\frac{d\phi}{dt}  &  =-\alpha\frac{d\theta}{dt}-\gamma\mu_{0}M_{s}N_{z}%
\theta\nonumber\\
\frac{d\theta}{dt}  &  =\alpha\frac{d\phi}{dt}+\gamma\mathcal{F}(I_{0})~,
\label{eq:bothmotion}%
\end{align}
introducing $\mathcal{F}(I_{0})=\hbar\Lambda_{st}I_{0}/(2eM_{s}V)$.
Eq.~(\ref{eq:bothmotion}) separates the motion for the in and out-of-plane
angles. We consider the dynamics of a current that is abruptly switched on to
a constant value $I_{0}$ at $t=0$, assuming that $\theta(t=0)=0$,
\textit{i.e.}, a magnetization that initially lies in the plane. The motion of
$\theta$ for $t>0$ is then given by
\begin{align}
\theta(t)  &  =\frac{\omega_{\phi}}{\gamma\mu_{0}M_{s}N_{z}}\left(
1-e^{-{t}/{\tau}}\right) \nonumber\\
\frac{d\theta}{dt}  &  =\frac{\alpha}{1+\alpha^{2}}\omega_{\phi}e^{-{t}/{\tau}}~.
\end{align}
where we introduced the response time
\begin{equation}
\tau=\frac{(1+\alpha^{2})}{\alpha\mu_{0}\gamma M_{s}N_{z}}~
\label{eq:resp-time}%
\end{equation}
and the saturation in-plane rotation frequency
\begin{equation}
\omega_{\phi}=\frac{\gamma\mathcal{F}(I_{0})}{\alpha}=\frac{\hbar}{2e}%
\Lambda_{st}\frac{\gamma I_{0}}{\alpha M_{s}V}~.
\end{equation}
Similarly, the in-plane rotation is governed by
\begin{align}
\phi(t)  &  =-\omega_{\phi}t+\frac{\omega_{\phi}}{\gamma\alpha\mu_{0}%
M_{s}N_{z}}\left(  1-e^{-{t}/{\tau}}\right) \nonumber\\
\frac{d\phi}{dt}  &  =-\omega_{\phi}+\frac{\omega_{\phi}}{1+\alpha^{2}%
}e^{-{t}/{\tau}}~.
\end{align}

Taking the parameters from Ref.~\onlinecite{vanwees}, \textit{viz}. a volume
of normal metal $V_{n}=400^{2}\times30~\text{nm}^{3}$, spin flip time in the
normal metal of $\tau_{sf}=62~\text{ps}$, density of states $\nu_{\text{D}
OS}=2.4\times10^{28}~\text{eV}^{-1}\text{m}^{-3}$, we find
$e^{2}g_{sf}/h=0.3~\Omega^{-1}$.

\begin{figure}[ptb]
\includegraphics[scale=0.45]{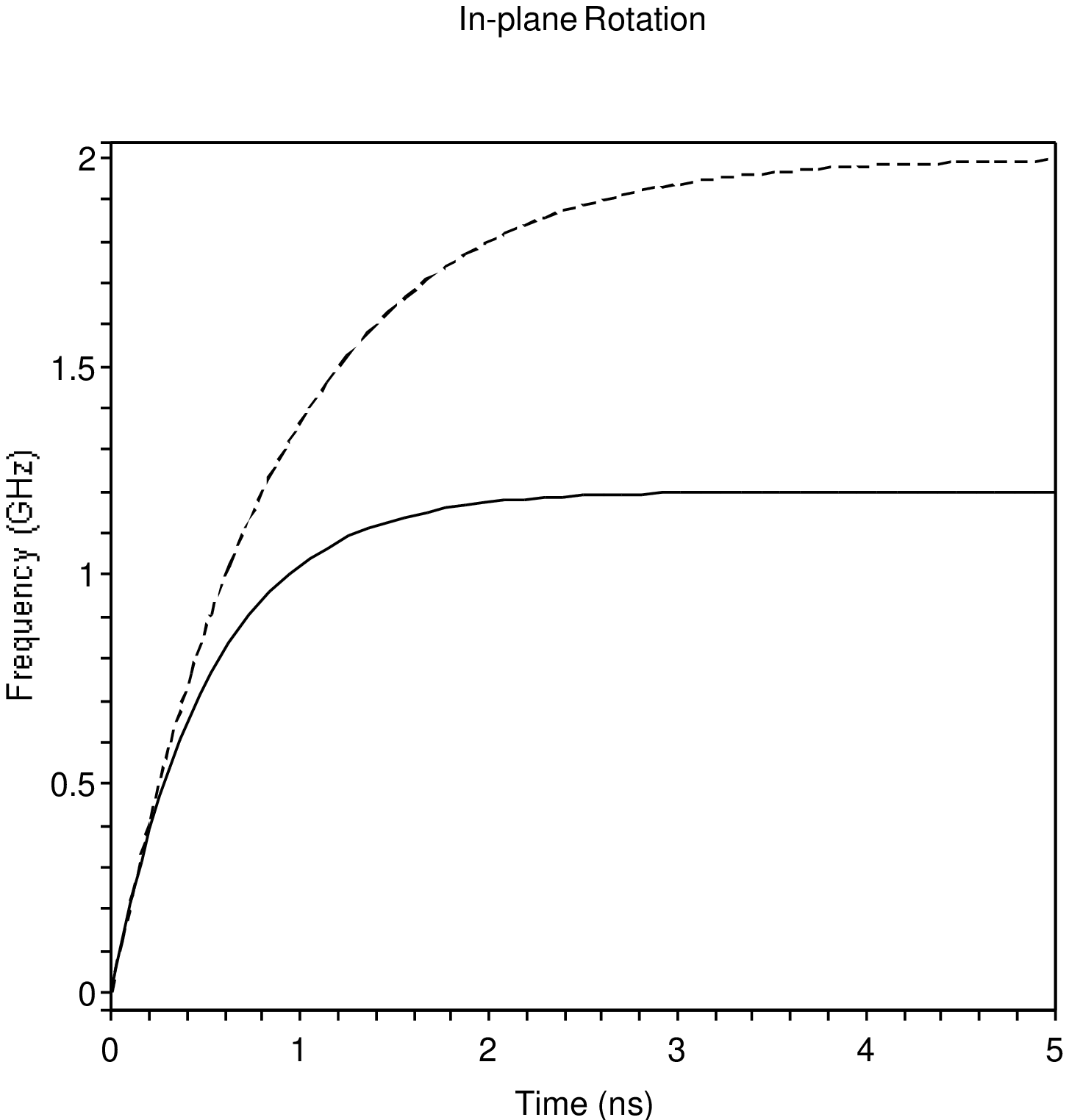}\caption{The in-plane rotation (in
the unit of giga hertz) versus time (in nano seconds). The solid line:
including spin pumping effect. The dash line: without spin spumping effect.}%
\label{fig:in-plane-velo}%
\end{figure}

\begin{figure}[ptbptb]
\includegraphics[scale=0.45]{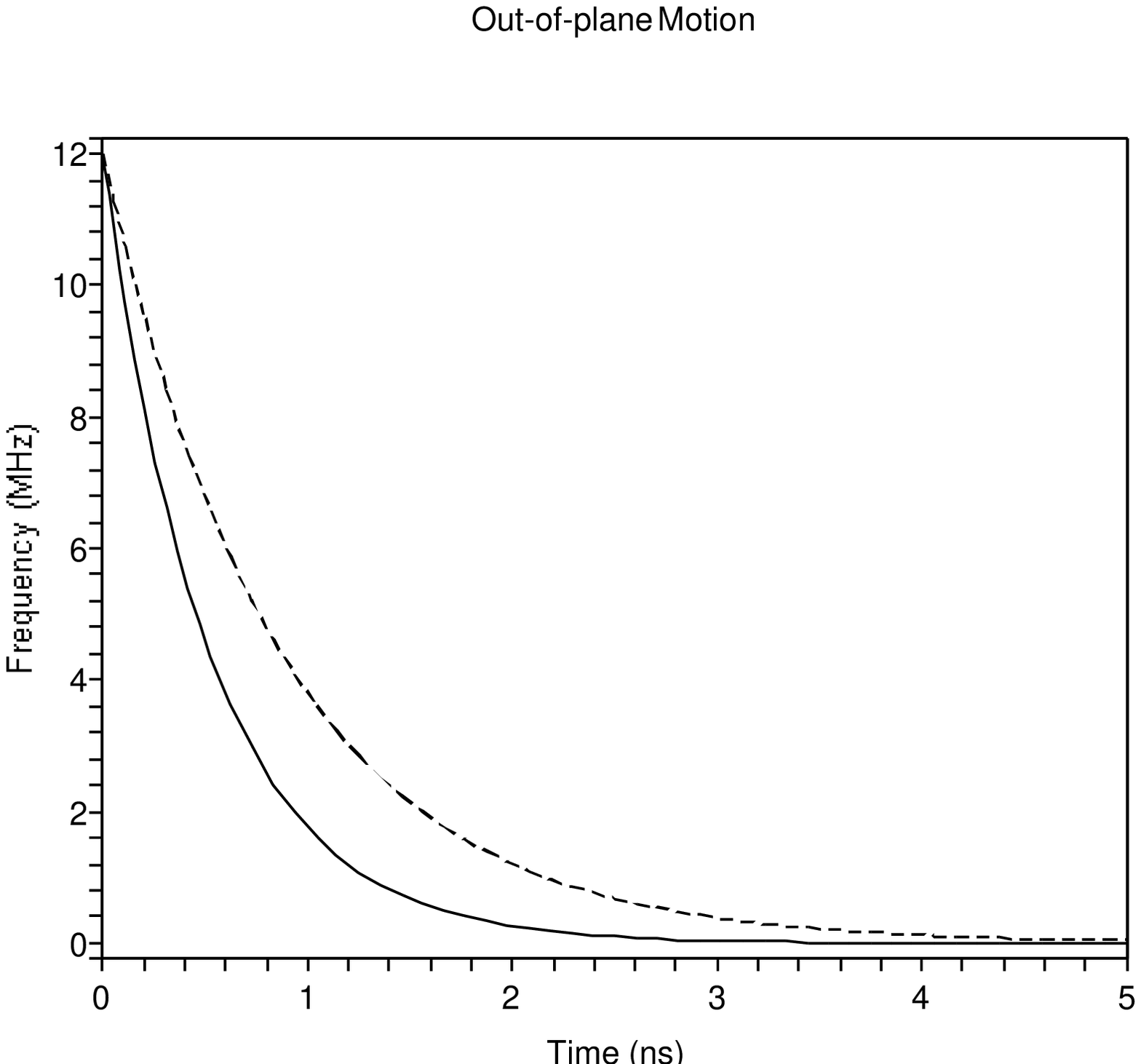}\caption{The out-of-plane
motion(in the unit of mega hertz) versus time (in nano seconds). The solid
line: including spin pumping effect. The dash line: without spin pumping
effect.}%
\label{fig:out-plane-velo}%
\end{figure}

Let us take the thickness of the free layer $d=5~\text{nm}$. The saturation
magnetization of permalloy is $M_{s}=8\times10^{5}~\text{A}~\text{m}^{-1}$.
The relative mixing conductance is chosen $\eta_{3}\simeq\eta_{r}\simeq1$ and
the bulk value of the Gilbert damping constant for Py is typically $\alpha
_{0}=0.006$.\cite{yaroslavprl} A metallic interface conductance (for $F3|N$)
is typically $1.3\text{f}\Omega~\text{m}^{2}$,\cite{kovalev}, whereas the
source/drain contacts are tunneling barriers with resistance $h/\left(
e^{2}g\right)  =20~\text{k}\Omega$.\cite{vanwees} The calculated enhancement
of the Gilbert damping constant is then $\alpha^{\prime}=0.004$ and the
response time $\tau=0.52~\text{ns}$. The motion of the magnetization of the
free layer is depicted by Fig.~\ref{fig:in-plane-velo} and
Fig.~\ref{fig:out-plane-velo} for a bias current density of $J=10^{7}
~\text{A~cm}^{-2}$ with the cross section at the electronic transport
direction $400\times30~\text{nm}^{2}$.~\cite{vanwees}

\begin{figure}[ptbptbptb]
\includegraphics[scale=0.45]{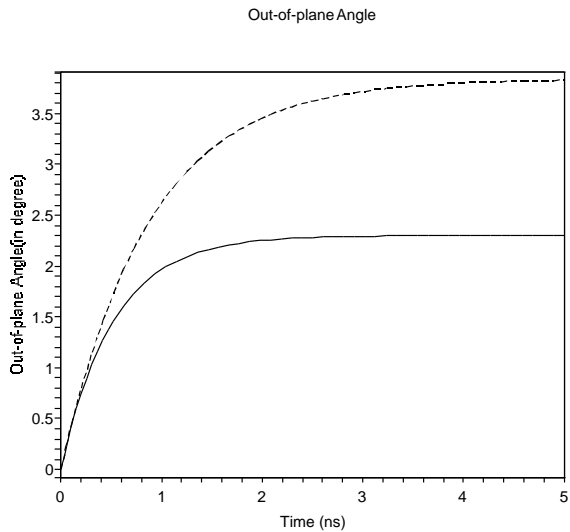}\caption{The out-plane angle
$\theta$ (in degree) versus time (in nano seconds). The solid line: including
spin pumping effect. The dash line: without spin spumping effect.}%
\label{fig:out-plane-angle}%
\end{figure}

The spin pumping effect through the enhanced Gilbert damping constant reduces
the saturation frequency from $2.0$ to $1.2~\text{GHz }$, but also the
response time to reach the saturation value from $0.87$ to $0.52~\text{ns}$.
Notice that the frequency is directly proportional to $I_{0}$ and thus in the
absence of any in-plane anisotropy the frequency can be tuned continuously to
zero by decreasing the bias current. The out-of-plane motion is very slow
compared to the in-plane one: it decreases from $12~\text{MHz}$ to around $0$
when the in-plane rotation approaches the saturation frequency. As shown in
Fig.~\ref{fig:out-plane-angle}, within a long period the small angle
approximation still holds. A larger ratio of $g_{3}/g$ also gives higher
frequencies. Decreasing the diameter, and thus also the volume, of the free
layer gives a smaller demagnetizing factor $N_{z}$, which causes larger a
response time $\tau$ according to Eq.~(\ref{eq:resp-time}) and increases the
saturation value of the in-plane rotation frequency $\omega_{\phi}$.



\subsubsection{In-plane anisotropy}

In reality, there are always residual anisotropies or pinning centers. Shape
anisotropies can be introduced intentionally by fabrication of elliptic F3
discs. We consider the situation in which the free layer is
slightly pinned in the plane by an anisotropy field that
corresponds to an elliptic (pancake) shape of the ferromagnet. At equilibrium,
the F3 magnetization is then aligned along the easy, let us say, $x-$axis. The
in-plane rotation can be sustained only when the spin transfer torque
overcomes the effective field generated by the shape anisotropy, hence a
critical current $I_{c}$ for the steady precession is expected. For an ellipse
with long axis of $200~\text{nm}$, thickness $5~\text{nm}$ and aspect ratio
$0.9$, the two demagnetizing factors are calculated to be $N_{y}=0.0224$ and
$N_{x}=0.0191$. With a Gilbert damping constant $\alpha=0.01$, the numerical
simulation gives $I_{c}=4.585~\text{mA}$ corresponding to a current density
$J_{c}=3.8\times10^{7}~\text{A}~\text{cm}^{-2}$ (the cross section is
$400\times30~\text{nm}^{2}$).~\cite{vanwees}

\begin{figure}[ptb]
\includegraphics[scale=0.43]{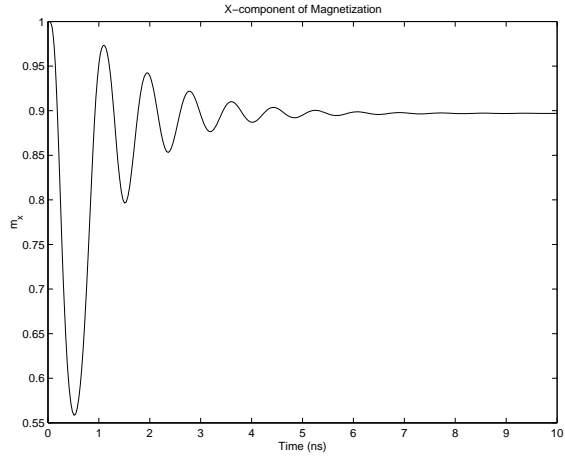}\caption{Below critical current, the
$x$-component of magnetization versus time (in nano seconds). The bias current
is $4.5~\text{mA}$.}%
\label{fig:bcx}%
\end{figure}

\begin{figure}[ptbptb]
\includegraphics[scale=0.43]{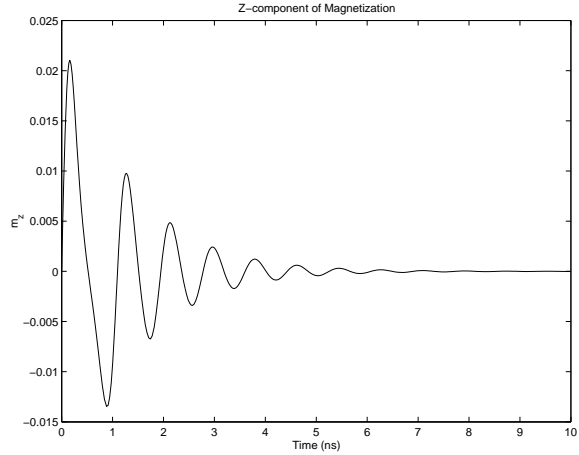}\caption{Below critical current, the
$z$-component of magnetization versus time (in nano seconds). The bias current
is $4.5~\text{mA}$.}%
\label{fig:bcz}%
\end{figure}

These critical current densities are of the same order of magnitude as those
used to excite the magnetization in spin-valve pillars. So even a relatively
small anisotropy can cause a significant critical current. In order to operate
the magnetic fan at small current densities, the magnetic island should be
fabricated as round as possible. The magnetization
responds to a current step function below the critical value by
damped in-plane and out-of-plane oscillations and
comes to rest at a new in-plane equilibrium angle $\phi_{e}$ with zero
out-plane component (\textit{cf}. Figs.~\ref{fig:bcx} and~\ref{fig:bcz}).
At the steady state, the spin-transfer torque is balanced by the
torque generated by the in-plane anisotropy, \textit{i.e.} the angle $\phi
_{e}$ is determined by $\sin(2\phi_{e})=2\mathcal{F}(I_{0})/(\mu_{0}M_{s}(N_{y}-N_{x}))$.
With given bias current, smaller $\left\vert N_{y}-N_{x}\right\vert $ correspond
to larger in-plane angles $\left\vert \phi_{e}\right\vert $. 
According to the theory of differential equations,\cite{perko}
the frequency for the damped magnetization oscillation can be found by diagonalizing the LLG
equation at the \textquotedblleft equilibrium point\textquotedblright\ given by $\phi_{e}$,
this leads to
\begin{equation}
\omega_{\phi}^{<}=\frac{\gamma\mu_{0}M_{s}}{\sqrt{2}}\sqrt{(2N_{z}%
-N_{x}-N_{y})\sqrt{\mathcal{D}(I_{0})}+\mathcal{D}(I_{0})}~,
\label{eq:fr-below-cc}
\end{equation}
where
\begin{equation}
\mathcal{D}(I_{0})=(N_{y}-N_{x})^{2}-\frac{4\mathcal{F}(I_{0})^{2}}{\mu
_{0}^{2}M_{s}^{2}}~.
\end{equation}
Equation (\ref{eq:fr-below-cc}) teaches us that
below the critical current, decreasing the current increases the rotation frequency.
Changing the damping constant does not change $\omega_{\phi}^{<}$ for a
given current but only changes the response time to reach the new equilibrium.

\begin{figure}[ptb]
\includegraphics[scale=0.43]{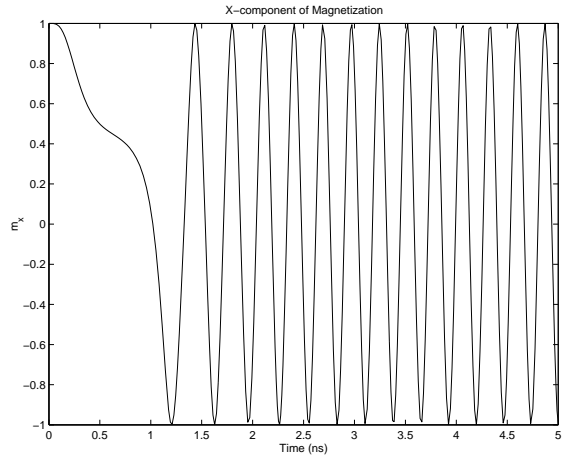}\caption{Above the critical current,
the $x$-component of magnetization versus time (in nano seconds). The bias
current is $4.6~\text{mA}$. The frequency is about $3.6$ GHz.}%
\label{fig:acx}%
\end{figure}

\begin{figure}[ptbptb]
\includegraphics[scale=0.43]{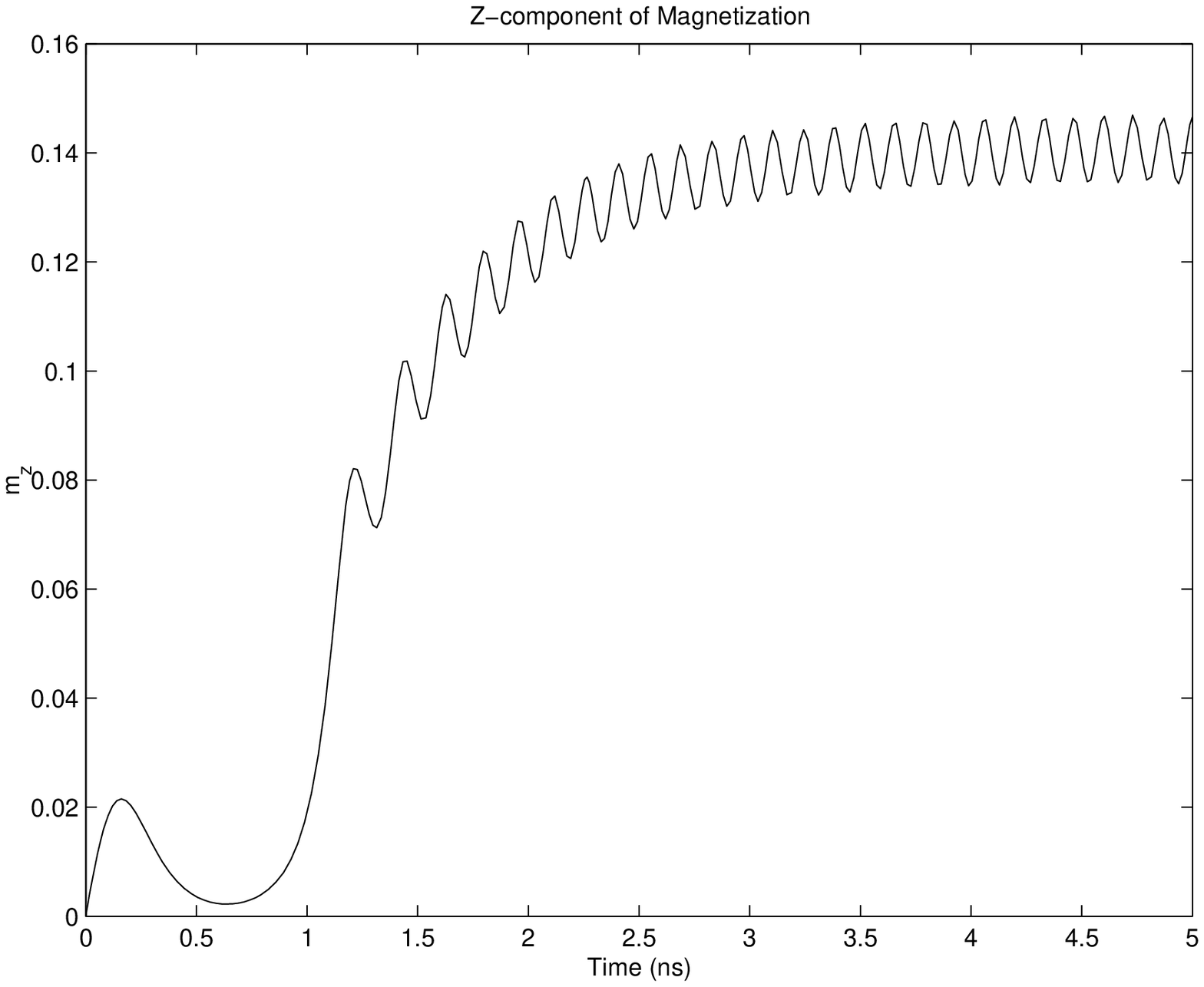}\caption{Above critical current, the
$z$-component of magnetization versus time (in nano seconds). The bias current
is $4.6~\text{mA}$.}%
\label{fig:acz}%
\end{figure}

\begin{figure}[ptbptbptb]
\includegraphics[scale=0.43]{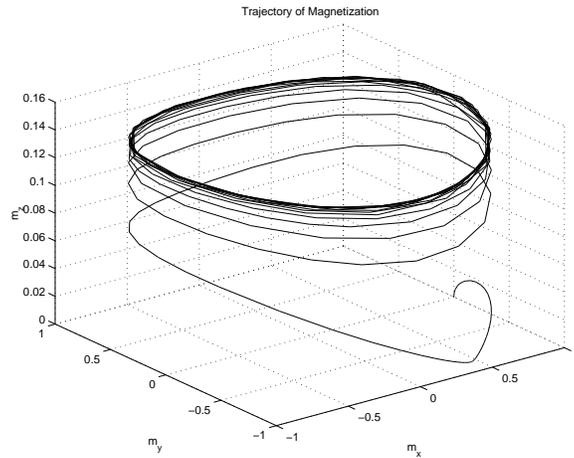}\caption{Above critical current, the
trajectory of magnetization within $5$ nano seconds. The bias current is
$4.6~\text{mA}$. This picture clearly shows the steady precession of the
magnetization.}%
\label{fig:acxyz}%
\end{figure}

\begin{figure}[ptbptbptbptb]
\includegraphics[scale=0.43]{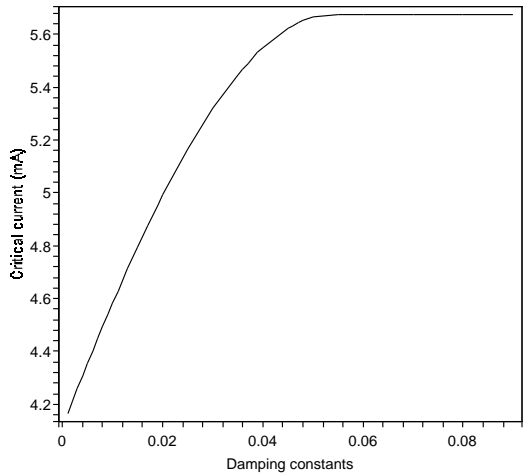}\caption{The critical current
$I_{c}$ versus damping constant $\alpha$. This figure shows saturation of
$I_{c}$ above a critical $\alpha$.}%
\label{fig:cc-damp}%
\end{figure}

As shown by Fig.~\ref{fig:acx} to Fig.~\ref{fig:acxyz} the magnetization above
the critical current saturates into a steady precessional state accompanied by
an oscillation of the $z$-component (nutation). In this situation, $\phi_{e}$
is no longer a constant of motion. Instead the new steady state is given
by $m_{x}=m_{y}=0$ and $\bar{m}_{z}=\mathcal{F}(I_{0})/(\alpha\mu_{0}
M_{s}N_{z})$. Diagonalizing the LLG around this point we derive the in-plane
rotation frequency
\begin{equation}
\omega_{\phi}^{>}=\frac{\gamma\mathcal{F}(I_{0})}{\alpha}\frac
{\sqrt{(N_{z}-N_{x})(N_{z}-N_{y})}}{N_{z}}~.
\end{equation}
In the limit of vanishing in-plane anisotropy, \textit{i.e.}, $N_{x}=0$ and
$N_{y}=0$, we recover the previous result. As shown by Fig.~\ref{fig:cc-damp}, the dependence of the critical
current on the damping constant is different from the simple proportionality
predicted for pillar structures.\cite{dieny} Specifically we observe
saturation of the critical current above a critical damping.

In the anisotropic case the extra power necessary for maintaining the motion
generates microwaves,\cite{kiselevnature,rippardprl} which may be attractive
for some applications.

\section{\label{sec:app}Applications}

Our \textquotedblleft magnetic fan\textquotedblright\ has the advantage that
the magnetization dynamics is not hidden within the structure as in the
pillars, but is open to either studies of the dynamics by fast microscopy, or
to the utilization of the dipolar field from the soft magnetic island. We
envisage applications as magnetic actuators for nanomechanical cantilevers and
nanoscale motors, as nanoscale mixers of biological or biomedical suspensions
containing magnetic nanoparticles, or as magnetic resonance detectors, again
possibly useful for biomedical applications.

\subsection{Actuators}

The rotating magnetization of the \textquotedblleft magnetic fan" generates a
periodic dipolar field which can be applied to actuate a nanomechanical
cantilever with a (hard) ferromagnetic tip. Assuming for simplicity that the
magnet F3 and the cantilever are at a sufficiently large distance the force on
the cantilever magnet is given by
\begin{equation}
\mathbf{F}=V_{c}\mathbf{\nabla}(\mathbf{M}_{c}\cdot\mathbf{H}_{d})~,
\end{equation}
where $\mathbf{M}_{c}$ is the saturation magnetization and $V_{c}$ is the
volume of the cantilever magnet and the field $\mathbf{H}_{d}$ generated by a
magnetic dipolar at the position $\mathbf{r}$ can be written as
\begin{equation}
\mathbf{H}_{d}=\mu_{0}\frac{3(\mathbf{M}\cdot\mathbf{r})\mathbf{r}%
-\mathbf{M}r^{2}}{r^{5}}~.
\end{equation}
Assume a cantilever on top of the magnetic fan at a distance of
$125~\text{nm}$ (along $z$-direction),\cite{rugarnature} with beam plane parallel
to the plane of the Py film F3 and magnetization along the $x$-axis. The saturation
value of cantilever magnetization is taken as
$1.27\times10^{6}~\text{A}~\text{m}^{-1}$. Assuming a lateral size of the cantilever
magnet~\cite{rugarnature} of $150\times150~\text{nm}^{2}$ with thickness
$50~\text{nm}$, the force is estimated to be
\begin{equation}
F=1.1\times10^{-8}\cos(\omega_{\phi}t)~\text{N}~
\end{equation}
where $\omega_{\phi}$ is the rotation frequency of the \textquotedblleft
magnetic fan". To efficiently generate the mechanical modes of the cantilever,
the cantilever magnet should be hard enough.

Fixing other parameters, the force scales like $1/r^{4}$ with respect to
distance $r$. When the two ferromagnets are closer to each other the
distribution of the magnetizations increases the force over the value
estimated above. We see that in the dipole-approximation, the force is already
quite significant and it will be significantly larger when the the full
magnetostatic energy is computed.

Generally, the torque on the cantilever may generate both flexural and
torsional motion on the cantilever. The torsional motion coupled to the
magnetization dynamics has been investigated for such a
system\cite{kovalevapl} and the nanomechanical magnetization reversal based on
the torsional modes has been proposed.\cite{kovalevprl} The coupling of a
cantilever to the oscillating dipolar field will be discussed elsewhere.

\subsection{Mixers}

The dipolar field produced by our device can also be used to function as
mechanical mixer for suspensions of magnetic particles. To this end we should
scale down the frequency of the rotating magnetization either by decreasing
the bias current or re-engineering the parameters of the device,
\textit{e.g.}, increasing the thickness of the Py film. Low saturation
magnetization is detrimental in this case, since that would also reduce the
usable stray fields. By these ways, one hopefully can access the kilo hertz
frequency region, which is important for the hydrodynamic motion in
ferrofluids.\cite{shliomis}

\subsection{Detectors}

An external field influences the frequency of the rotation of the
magnetization. Response to the change of the frequencies is the rebuilding of
the spin accumulation in the normal metal hence altering the source-drain
resistance $R_{SD}$. Due to the relation
\begin{equation}
\mu_{0}^{F1}-\mu_{0}^{F2}=R_{SD}I_{0}~,
\end{equation}
this deviation is reflected on the source-drain voltage-current curve. This
feature can be implemented as a sensor for biomedical applications in order to
detect the presence of magnetic beads, which are used as labels in
biosensors.\cite{BARC} Furthermore, the ability to change the frequency of the
\textquotedblleft magnetic fan" should allow to measure locally the frequency
dependence of the magnetic susceptibility, which offers an alternative pathway
to using magnetic nanoparticles for biosensing
applications.\cite{stpierre,chung}

\section{\label{sec:concl}Conclusion}

We studied the magnetization dynamics of a magnetic transistor, \textit{i.e.},
a lateral spin valve structure with perpendicular-to-plane magnetizations and
an in-plane free layer attached to the normal metal that is excited by an
external current bias. By circuit theory and the Landau-Lifshitz-Gilbert
equation, analytic results were obtained for the spin-transfer torque and the
dynamics of the magnetization in the limit of small out-of-plane angle
$\theta$. Spin flip and spin-pumping effects were also investigated
analytically and an anisotropic enhanced Gilbert damping was derived for the
free layer magnetization. Without an externally applied magnetic field, a
continuous rotation of the magnetization of the free layer at GHz frequencies
can be achieved. In the lateral geometry, the free layer is no longer buried
or penetrated by a dissipating charge current, thus becomes accessible for more
applications. Our methods handle the microscopic details on crucial
issues like spin-torque transfer efficiency, spin-flip scattering and the
response time, hence offering accurate design and control. The rotation can be
observed, \textit{e.g.}, by magneto-optic methods. This new device has
potential applications as high frequency generator, actuator of nanomechanical
systems, biosensors, and other high-speed magnetoelectronic devices.

\begin{acknowledgments}
We thank Yu. V. Nazarov, J. Slonczewski and A. Kovalev 
for fruitful discussions. X. Wang acknowledges H.
Saarikoski's help with the preparation of the manuscript. This work is
supported by NanoNed, the FOM, U.S.\ Department of Energy, Basic
Energy Sciences, under Contract No.~W-31-109-ENG-38 and the EU
Commission FP6 NMP-3 project 505587-1 \textquotedblleft
SFINX\textquotedblright. GEWB would like to acknowledge the support
through the Materials Theory Institute and the hospitality he
enjoyed at Argonne.
\end{acknowledgments}

\begin{appendix}
\section{\label{sec:appa}Spin accumulation in normal metal node}
Here we summarize a number of complex angle dependent coefficients.
The elements of the symmetric matrix $\hat{\mathbf{C}}$ in
Eq.(\ref{eq:spin-accu}) read
\begin{widetext}
\begin{align}
C_{11}=&\frac{\mathcal{G}_{t}(\mathcal{G}_{1}-\mathcal{G}_{4}m_{x}^{2})-2g(p^{2}-1+\eta)(\mathcal{G}_{t}-\mathcal{G}_{4}m_{y}^{2})}{\mathcal{Q}}\\
C_{12}=&C_{21}=\frac{\mathcal{G}_{2}\mathcal{G}_{4}m_{x}m_{y}}{\mathcal{Q}},\quad\text{and}\quad
C_{13}=C_{31}=\frac{\mathcal{G}_{t}\mathcal{G}_{4}m_{x}m_{z}}{\mathcal{Q}}\\
C_{22}=&\frac{\mathcal{G}_{2}(\mathcal{G}_{t}-\mathcal{G}_{4}m_{x}^{2})-\mathcal{G}_{t}\mathcal{G}_{4}m_{z}^{2}}{\mathcal{Q}}\\
C_{23}=&C_{32}=\frac{\mathcal{G}_{t}\mathcal{G}_{4}m_{y}m_{z}}{\mathcal{Q}},\quad\text{and}\quad
C_{33}=\frac{\mathcal{G}_{t}(\mathcal{G}_{1}+\mathcal{G}_{4}m_{z}^{2})}{\mathcal{Q}}
\end{align}
\end{widetext}
introducing
\begin{widetext}
\begin{align}
\mathcal{G}_{1}& =(1-p_{3}^{2})(1-\Delta_{3})g_{3}+2\eta g + 2g_{sf}\\
\mathcal{G}_{2}& =\eta_{3}g_{3}+2(1-p^{2})g + 2g_{sf}\\
\mathcal{G}_{3}& =(1-p_{3}^{2})(1-\Delta_{3})g_{3}+2(1-p^{2}) g + 2g_{sf}\\
\mathcal{G}_{4}& =\eta_{3}g_{3}-(1-p_{3}^{2})(1-\Delta_{3})g_{3}\\
\mathcal{G}_{t}& =\eta_{3}g_{3}+2\eta g+2g_{sf}\\
\mathcal{Q}&
=\mathcal{G}_{t}[\mathcal{G}_{t}\mathcal{G}_{3}+2(p^{2}-1+\eta)g\mathcal{G}_{4}(1-m_{z}^{2})]\\
\Delta_{3}& =\frac{\zeta_{3}}{\zeta_{3}+\tilde{\sigma}\tanh(d/l_{sd}^{F})}~,
\end{align}
\end{widetext}
in the limit of negligible spin flip in F, i.e., $d\ll l_{sd}^{F}$, then $\Delta_{3}\approx 1$.
The elements of the matrix in Eq.(\ref{eq:spin-torque}) are given by
\begin{widetext}
\begin{align}
\Pi_{11}=&
\frac{\mathcal{G}_{t}\mathcal{G}_{3}(1-m_{x}^{2})+2\mathcal{G}_{4}(p^{2}-1+\eta)gm_{y}^{2}}{\mathcal{Q}}\\
\Pi_{12}=&\Pi_{21}=\frac{-\mathcal{G}_{1}\mathcal{G}_{2}m_{x}m_{y}}{\mathcal{Q}},\quad\text{and}\quad
\Pi_{13}=\frac{-\mathcal{G}_{t}\mathcal{G}_{1}m_{x}m_{z}}{\mathcal{Q}}\\
\Pi_{22}=&
\frac{\mathcal{G}_{t}\mathcal{G}_{3}(1-m_{y}^{2})+2\mathcal{G}_{4}(p^{2}-1+\eta)gm_{x}^{2}}{\mathcal{Q}}\\
\Pi_{23}=&\frac{-\mathcal{G}_{t}\mathcal{G}_{1}m_{y}m_{z}}{\mathcal{Q}},\quad\text{and}\quad
\Pi_{31}=\frac{-\mathcal{G}_{t}\mathcal{G}_{3}m_{x}m_{z}}{\mathcal{Q}}\\
\Pi_{32}=&\frac{-\mathcal{G}_{t}\mathcal{G}_{3}m_{y}m_{z}}{\mathcal{Q}},\quad\text{and}\quad
\Pi_{33}=\frac{\mathcal{G}_{t}\mathcal{G}_{1}(1-m_{z}^{2})}{\mathcal{Q}}\\
\end{align}
\end{widetext}

\end{appendix}

\bibliography{propeller}

\end{document}